\documentclass[aps,prb,10pt,twocolumn,superscriptaddress,longbibliography]{revtex4-2}
\usepackage{graphicx} 
\RequirePackage{lineno}
\usepackage[utf8]{inputenc}
\usepackage[a4paper, margin=0.5in]{geometry}
\usepackage{graphicx} 
\usepackage{caption}
\usepackage{subcaption}
\usepackage{hyperref}
\usepackage{amsthm}
\usepackage{enumitem}
\usepackage{braket}
\usepackage{color}
\usepackage{amssymb,amsmath,amsfonts,gensymb}
\usepackage{ulem}
\usepackage{dcolumn}
\usepackage{mathtools}
\usepackage{pifont}
\usepackage{bm,hyperref}
\usepackage{cleveref}

\newcommand{\mc}[1]{\mathcal{#1}}
\newcommand{\mr}[1]{\mathrm{#1}}

\def\k{\mathbf{k}}
\def\nn{\nonumber \\}
\begin{document}
\title{Anomalous Chern-Simons orbital magnetoelectric coupling of three-dimensional Chern insulators: gauge-discontinuity formalism and adiabatic pumping}
\author{Yang Xue} 
\affiliation{School of Physical Science and Technology, ShanghaiTech University}
\author{Jianpeng Liu}
\email{liujp@shanghaitech.edu.cn}
\affiliation{School of Physical Science and Technology, ShanghaiTech University}
\affiliation{ShanghaiTech Laboratory for Topological Physics, ShanghaiTech University}
\affiliation{Liaoning Academy of Materials, Shenyang 110167, China}

\begin{abstract}
Chern-Simons orbital magnetoelectric (OME) coupling is usually the hallmark of nontrivial band topology in three-dimensional (3D) crystalline insulators. However, if a 3D insulator exhibits nonzero Chern number within any two-dimensional plane of the Brillouin zone, then traditionally the Chern-Simons coupling becomes ill defined for such 3D Chern insulators due to topological obstructions. In this work, by employing a ``gauge-discontinuity" formalism, we resolve this long-standing issue and rigorously derive a quantized layer-resolved OME response in 3D Chern insulators. We demonstrate that the difference of the layer-resolved OME coupling between adjacent layers is universally quantized in unit of $-C e^2/h$, where $C$ is the Chern number. This quantization arises from an anomalous contribution to the Chern-Simons OME coupling, which is closely associated with the Berry curvature of the occupied bands and the hybrid Wannier centers along the direction of the Chern vector $(0,0, C)$. Furthermore, we demonstrate that the anomalous Chern-Simons coupling can be transported by an exact integer quantum from one unit cell to its neighboring cell through an adiabatic cyclic pumping process, accompanied by a quantized displacement of Wannier center along the direction of the Chern vector.
Our work provides a rigorous theoretical framework for understanding magnetoelectric response in 3D Chern insulators and opens avenues for designing topological quantum phenomena in layered systems.
\end{abstract}

\maketitle
\section{Introduction}
Magnetoelectric coupling is an effect in some insulating solids where an applied electric field $\mathcal{\mathbf{E}}$ can induce magnetization $\mathbf{M}$, and reversely, a magnetic field $\mathbf{B}$ can generate electric polarizaton $\mathbf{P}$. The linear magnetoelectric coupling coefficient can be properly described using a rank-2 tensor, 
\begin{equation}
\alpha_{a,b}=\frac{\partial M_b}{\partial E_a}{\Big\vert}_{\mathbf{E}=0}
=\frac{\partial P_a}{\partial B_b}\Big\vert_{\mathbf{B}=0}\;,
\end{equation}
where $a, b= x, y, z$ are spatial coordinates.
Depending on the origin of the 
magnetization induced by electric field, magnetoelectric coupling may have both spin and orbital contributions \cite{malashevich2010ome,fiebig2005review}. For the spin component of ME coupling, external electric field would cause the change of spin magnetizations of electrons through various complex mechanisms including spin-orbit coupling and spin-lattice coupling etc., which has been extensively discussed and studied in materials with co-existing magnetic and electric order parameters such as multiferroics \cite{tokura-2014review,fiebig-nrm2016,spaldin-nm2019}. The orbital contribution to magnetoelectric coupling, on the other hand, is more subtle and intriguing, which is usually closely related to the geometric and topological properties of the Bloch states of solids \cite{malashevich2010ome,essin-prb10}. An typical example is the Chern-Simons orbital magnetoelectric coupling denoted as $\alpha^{\rm{CS}}$, a diagonal and isotropic orbital magnetoelectric response \cite{qi-prb08,essin-prl09,malashevich2010ome,essin-prb10}, 
\begin{equation}
\alpha_{a,b}^{\rm{CS}}=\frac{\theta}{2\pi}\frac{e^2}{h}\,\delta_{ab}\;,
\end{equation}
which is characterized by a dimensionless phase angle $\theta$, and is gauge invariant modulo $2\pi$.

Clearly, magnetoelectric coupling (including the Chern-Simons coupling) is odd under both time-reversal ($\mathcal{T}$) and inversion ($\mathcal{P}$) operations. Therefore, for a system with either $\mathcal{T}$ or $\mathcal{P}$ symmetry, $\theta$ has to 0 or $\pi$ (thanks to the $2\pi$ ambiguity), which imposes a $\mathbb{Z}_2$ classification to $\mathcal{T}$-invariant and $\mathcal{P}$-invariant crystalline insulators \cite{qi-prb08,essin-prl09,hughes-prb11,turner-prb12}. Those 3D insulators exhibiting half-quantized bulk Chern-Simons coupling $\theta=\pi$ ($\alpha^{\rm{CS}}=e^2/2h$) are known as $\mathcal{T}$-invariant topological insulators \cite{kane-rmp10,zhang-rmp11} or $\mathcal{P}$-invariant axion insulators \cite{hughes-prb11,turner-prb12}. However, so far the Chern-Simons OME coupling is well defined only for 3D bulk insulators with zero Chern number for every two-dimensional cut of the Brillouin zone. This is because $\theta$ can be expressed as an integral of the Chern-Simons 3-form defined by the occupied Bloch states over the 3D Brillouin zone, which necessarily requires the presence of a smooth and 
periodic gauge for the Bloch states throughout the Brillouin zone. Such a gauge is non-existing if the Chern number of occupied states is nonzero within some 2D plane in reciprocal space \cite{exponential-wannier-prl07}. 
Such a 3D charge insulating system with nonzero Chern number (defined within certain 2D plane of Brillouin zone) is known as 3D Chern insulator, which would exhibit three-dimensional quantized anomalous Hall effect, and is adiabatically connected to an infinite number of vertically stacked, mutually decoupled 2D Chern-insulator layers. 
The Chern-Simons orbital magnetoelectric coupling for 3D Chern insulators has been an unsolved theoretical problem due to the topological obstructions imposed by Chern number. 

Recently, it has been shown by extensive numerical simulations and semi-analytical arguments that in a 3D Chern insulator consisting of vertically stacked 2D Chern layers, the ``layer-resolved'' orbital magnetoelectric coupling obeys an integer quantization rule: $\alpha_{l+1}-\alpha_{l}=-C e^2/h$ \cite{lu_2024}, where $\alpha_l$ denotes the orbital magnetoelectric coupling coefficient projected to layer $l$, $C$ is the layer Chern number, $e$ is elementary charge, and $h$ is Planck constant. Such a quantization rule seems to remain exact in the presence of various types of interlayer couplings, stackings, and in the presence of substrate effects and disorder \cite{lu_2024}. This motivates us to theoretically investigate the origin of such quantized difference of layer-resolved orbital magnetoelectric response. In this work, by adopting the ``gauge-discontinuity'' formalism of Chern-Simons OME coupling \cite{Liu-GD-2015}, we manage to prove the above quantization rule in an mathematically exact manner for generic 3D Chern insulators. We will show that such a layer-quantized magnetoelectric coupling may be interpreted as an ``anomalous'' Chern-Simons coupling, where the ``anomaly'' is precisely from the nonzero Chern number.

Moreover, experimentally the bulk Chern-Simons magnetoelectric coupling is indistinguishable from a surface anomalous Hall effect \cite{essin-prl09,essin-prb10}. Namely, an applied electric field would induce anomalous Hall currents at all surfaces, and these surface anomalous Hall currents would wind around the sample, thus contribute an orbital magnetization in the direction of electric field. This also naturally explains the origin of the $2\pi$ ambiguity of the Chern-Simons coupling. In principle, one can always coat a Chern-insulator layer with Chern number $C$ to all surfaces of a crystalline insulator, such that the surface anomalous Hall effect involves contributions both from the bulk Chern-Simons coupling $(\theta/2\pi)e^2/h$ and the surface Chern layer $C e^2/h$. The extra integer contribution from surface Chern layers cannot be uniquely determined by bulk properties, therefore the bulk Chern-Simons coupling $\theta$ has an ambiguity of $2\pi$ \cite{qi-prb08,essin-prl09}. This is in close analogy with the integer ambiguity of bulk electric polarization of one-dimensional band insulators \cite{ks-vanderbilt-prb93}, which originates from the non-determined end charges \cite{vanderbilt-ks-prb93}. The integer ambiguity of electric polarization allows one to design an cyclic adiabatic process, in which an integer number of charges are pumped from one end to the other while leaving the bulk state unchanged \cite{thouless-prb83}. Furthermore, it has been theoretically proposed that the Chern-Simons coupling of $\mathcal{T}$-invariant 3D insulators with zero Chern number, when expressed in proper hybrid Wannier basis, can be pumped by an integer quantity $e^2/h$ through a $\mathcal{T}$-breaking adiabatic evolution \cite{Maryam-2015}. Here, we will show that the ``anomalous" Chern-Simons coupling of a 3D Chern insulator (expressed in proper hybrid Wannier basis) can be pumped by an integer quantized value $-C e^2/h$ from one unit cell to the neighboring cell via a different mechanism: in our case of 3D Chern insulators, such quantized pumping of $\theta$ can be realized with single occupied band, and is always concomitant with the integer pumping of charge.

The remaining part of the paper is organized as follows. In Sec.~\ref{sec:preliminary} we will briefly review the mathematical expression of Chern-Simons OME coupling and the gauge-discontinuity formalism in treating it. In Sec.~\ref{sec:formalism}, we extend the gauge-discontinuity formalism of Chern-Simons coupling to the case of 3D Chern insulators with single occupied band, and derive an ``anomalous" contribution to the Chern-Simons coupling arising due to the nonzero Chern number. In Sec.~\ref{sec:multi-band}, we extend the single-band formalism of Chern-Simons coupling of 3D Chern insulators to the case of multiple occupied bands. In Sec.~\ref{sec:quantization}, we explicitly demonstrate how an exactly quantized difference of layer-projected Chern-Simons OME coupling would emerge from the above formalism. In Sec.~\ref{sec:pumping}, we devise an cyclic adiabatic path for a multilayer extension of Haldane model, through which the layer-projected Chern-Simons OME coupling is pumped by exactly $-C e^2/h$ from one unit cell to its neighboring cell. Finally, in Sec.~\ref{sec:summary} we make a summary.

\section{Preliminary}
\label{sec:preliminary}
\subsection{Review of Chern-Simons OME coupling}
The topological orbital magnetoelectric coupling for insulators with vanishing Chern number is given by the integral of the Chern-Simons 3-form \cite{malashevich2010ome,essin-prb10,essin-prl09,qi-prb08}
\begin{equation}
\theta_{\text{CS}}=-\frac{1}{4\pi} \int d^3k\,\epsilon_{abc}\mr{Tr}[A_a\partial_b A_c-\frac{2i}{3}A_a A_b A_c]
\label{eq:integral of CS}
\end{equation}
where $A_{a,mn}=\braket{u_{m\k}|\partial_{_a}|u_{n\k}}$ is the Berry connection matrix element of the occupied Bloch states at wavevector $\k$, and the notation $\partial_a\equiv \partial_{k_a}$ with $a=x,y,z$. $\vert u_{n\k} \rangle$ ($\vert u_{m\k} \rangle$) denotes the periodic part of the Bloch state at $\k$ with band index $n$ ($m$). Mathematically, the integral of the three form's exterior derivative over a 4-dimensional compact manifold, which can be interpreted as the ``four-dimensional Brillouin zone" of a 3D insulator under a cyclic adiabatic parameter $\eta$,
\begin{equation}
C^{(2)}\equiv\frac{1}{32\pi^2}\int d^3k\,d\eta\,\epsilon^{abcd}\mr{Tr}(F_{ab}F_{cd})\;,
\end{equation}
is the second Chern number, with $a, b, c,d=x,y,z,\eta$. The generalized Stokes' theorem would thus show that a non-trivial pumping of $\theta$ through a cyclic adiabatic evolution is dictated by a non-zero second Chern number. 

Nevertheless, the Chern-Simons magnetoelectric coupling $\theta$ (Eq.~\eqref{eq:integral of CS}) is well studied only for 3D insulators with vanishing Chern numbers in every 2D cut of the 3D Brillouin zone. Several alternative representations of Chern-Simons coupling $\theta$ have been proposed. For example, it has been shown in \cite{Maryam-2015} that $\theta$ can be expressed in the hybrid Wannier representation, 
\begin{align}
\theta=&-\frac{1}{d_0}\int d^2k\sum _n\bar{z}_n\,\Omega_{xy,0n,0n}\nn&+\frac{i}{d_0}\int d^2k\sum_{lmn}(\bar{z}_{lm}-\bar{z}_{0n})A_{x,0n,lm}A_{lm,0n}
\label{triv-wannier-repre}
\end{align}
where $\Omega_{xy,0n,0n}$, $A_{a,0n,lm}$ are the Berry curvature and Berry connection matrix elements defined in the hybrid Wannier basis, $\bar{z}_n$ is the $n$th hybrid Wannier center, and $d_0$ is the lattice constant in $z$ direction. This draws the connection between Berry curvature (Berry connection), Wannier center and Chern-Simons coupling $\theta$. However, still it only applies to insulators with vanishing Chern numbers. 

Since Chern-Simons 3-form involves Berry connection \(A_a\), it becomes ill-defined at some points in the 3D Brillouin zone when a global smooth and periodic gauge of the occupied Bloch states is missing. This is exactly the case for 3D Chern insulators, where the Chern number of a given 2D $\k$ plane (say, the $(k_x,k_y)$ plane) is a nonzero integer $C$. The topological obstruction imposed by nonzero Chern number necessarily introduces singularities or discontinuities of the gauge of occupied Bloch states somewhere in the Brillouin zone.

Therefore, in order to study the Chern-Simons OME coupling of 3D Chern insulators, we are forced to consider the contribution of ``gauge discontinuity" somewhere in the 3D Brillouin zone. In Ref.~\onlinecite{Liu-GD-2015}, a new formalism to treat the Chern-Simons OME coupling has been proposed. For example, in the case of 3D topological insulators, nontrivial $\mathbb{Z}_2$ band topology protected by $\mathcal{T}$ symmetry would cause obstructions in constructing a smooth and periodic gauge respecting $\mathcal{T}$ symmetry, making it hard to numerically achieve a half quantized value $\theta=\pi$. To solve this problem, the authors of Ref.~\onlinecite{Liu-GD-2015} proposed to retain the smoothness of a $\mathcal{T}$-preserving gauge, while relaxing the periodicity condition, which introduces some ``gauge discontinuity" at certain 2D boundary of the 3D Brillouin zone. Then, the contributions from the gauge-discontinuity plane is further taken into account in addition to the bulk integral of Chern-Simons 3-form. Since the periodicity condition has been relaxed, the gauge can be made as smooth as possible in the bulk of 3D Brillouin zone, which makes the numerical integral of the bulk Chern-Simons 3-form much easier. Indeed, in the case of 3D topological insulators, it has been shown that the half-quantized $\theta=\pi$ is all contributed by one particular term originating from the gauge-discontinuity plane (known as ``vortex loop" term, to be elaborated later) \cite{Liu-GD-2015}, and the bulk integral of Chern-Simons 3-form vanishes due to the choice of a smooth and $\mathcal{T}$-preserving gauge. 

In the case of 3D Chern insulators, we are facing a similar situation: one cannot make a smooth and periodic gauge (even for single occupid band) throughout the Brillouin zone due to topological obstructions imposed by nonzero Chern number. Therefore, we would like to make the gauge as smooth as possible in the bulk of 3D Brillouin zone, while relaxing the periodicity condition of the gauge at a certain 2D boundary. Then, we will apply the gauge-discontinuity formalism of Chern-Simons 3-form to the case of 3D Chern insulators, and consider the bulk and gauge-discontinuity contributions separately.

\subsection{Review of gauge discontinuity formalism}
\begin{figure*}[hbt]
  \centering
  \includegraphics[width=\textwidth]{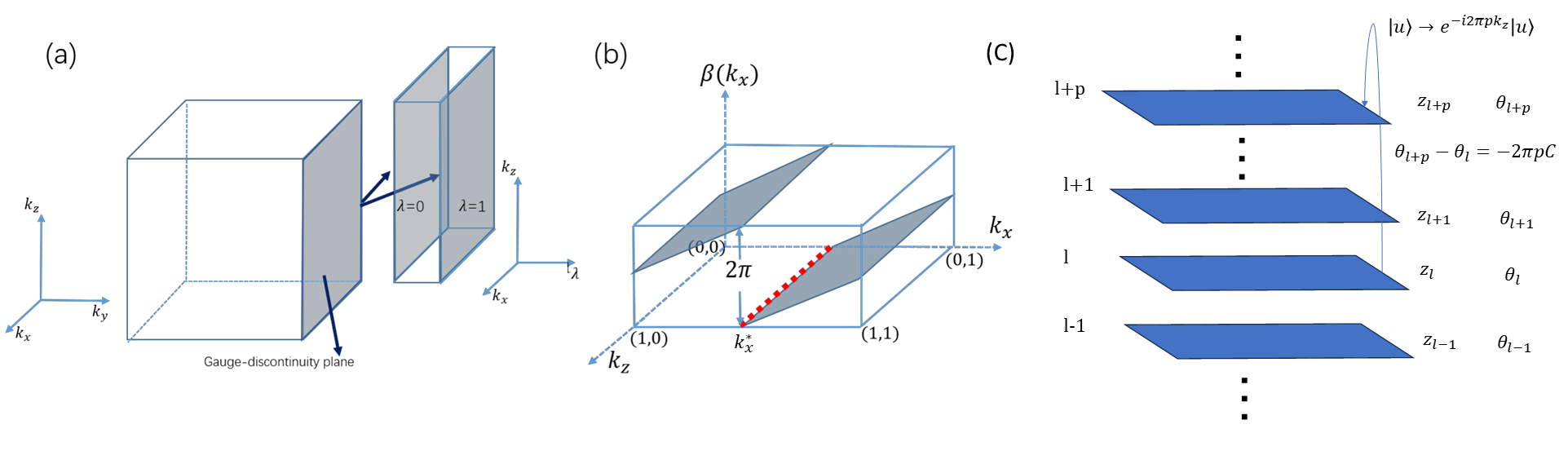}
  \caption{(a) An illustration of a cylinder gauge with a discontinuity at $k_y=0$ being connected by a $\lambda$-space. (b) An illustration of the discontinuity characterized by $\beta(k_z,k_x)$ on the $k_y=0$ plane. (c) The gauge transformation transforming $\theta$ of layer $l$ to that of layer $l+p$}
  \label{Fig: formalism}
\end{figure*}

In this subsection, we briefly review the gauge-discontinuity formalism introduced in Ref.~\onlinecite{Liu-GD-2015}. We first specify a gauge that is convenient for our analysis. Throughout the paper we use reduced coordinates $k_x$, $k_y$, $k_z\in[0,1)$. Consider a gauge with discontinuity at$k_y=1$ but smooth along $k_z$ and $k_x$. 
\begin{align}
\ket{\psi_n(k_x,k_y=1,k_z)}=&\sum_{m=1}^M W_{mn}(k_x,k_z)\nn
&\times\ket{\psi_m(k_x,k_y=0,k_z)}
\end{align}
where $W=e^{iB}$ is an $M\times M$ unitary matrix defined in the basis of the $M$ occupied Bloch states $\{\ket{\psi_n(k_x,k_y,k_z)}\}$, characterizing the gauge discontinuity across the $k_y=0, 1$ plane and $B$ is a Hermitian matrix. To handle the discontinuity, the formalism introduces a fictitious \(\lambda\)-space, in which we linearly interpolate the discontinuity from $k_y=1$ to $k_y=0$, 
\begin{equation}
\ket{\psi_n(k_x,\lambda,k_z)}=\sum_{m=1}^M W_{mn}\ket{\psi_m(k_x,\lambda=0,k_z)}
\label{eq:gauge-disc}
\end{equation}
where
\begin{equation}
  \ket{\psi_n(k_x,\lambda=0,k_z)}=\ket{\psi_n(k_x,k_y=1,k_z)}
\end{equation}
In the $\lambda$-space, the Berry connection and Berry curvature are defined as usual with respect to $\ket{u(k_x,\lambda,k_z)}$. With above definition, we have
\begin{subequations}
\begin{align} 
&\ket{\psi_n(k_x,\lambda=1,k_z)}=\ket{\psi_n(k_x,k_y=0,k_z)}\\
&\ket{u(k_x,\lambda=1,k_z)}=e^{-i2\pi \hat{y}}\ket{u(k_x,k_y=0,k_z)}
\end{align}
\end{subequations}
Namely, we have removed the discontinuity by smoothly connecting the $k_y=0$ and $k_y=1$ planes with a fictitious $\lambda$ space, as schematically shown in Fig.~\ref{Fig: formalism}(a), 

The Chern-Simons OME thus has three contributions
\begin{equation}
  \theta=\theta_{\text{BK}}+\theta_{\text{GD}}+\theta_{\text{VL}}
\label{eq:three-thetas}
\end{equation}
where $\theta_{\text{BK}}$ is the contribution from the integral of Chern-Simons 3-form within the bulk Brillouin zone, but without enforcing periodicity at the $k_y=0, 1$ plane. $\theta_{\text{GD}}$ and $\theta_{\text{VL}}$ are referred as the ``gauge-discontinuity" and ``vortex-loop" terms both arising from the contribution of the fictitious $\lambda$ space connecting $k_y=0$ and $k_y=1$ planes.
Following directly from the Eqs.~\eqref{eq:integral of CS}, after some integrations by part, the first two terms in Eq.~\eqref{eq:three-thetas} are given by
\begin{subequations}
\begin{align}
\theta_{\textrm{BK}} =& -\frac{1}{4\pi}\int d^3k\,\mr{Tr} [A_x\Omega_{yz} \nn
& + A_y\Omega_{zx} + A_z\Omega_{xy} - 2i [A_x, A_y] A_z] \label{eq:pre-thetaBulk} \\
\theta_{\text{GD}} =& -\frac{1}{4\pi}\int dk_z\,dk_x\,d\lambda\,\mr{Tr} [A_z\Omega_{x\lambda} \nn
& + A_x\Omega_{\lambda z} + A_\lambda\Omega_{zx} - 2i [A_z, A_x] A_\lambda] \label{eq:pre-thetaGD}
\end{align}
\end{subequations}
where the trace is carried out over the occupied bands. We note that the unitary gauge discontinuity matrix $W=e^{i B}$ can be equivalently described by the Hermitian matrix $B$ in its exponent. The eigenvalues of $B$ thus have $2\pi$ ambiguities. Sometimes, it is impossible to insist a branch choice such that all eigenvalues of $B$ would remain smooth in $k_z$ and $k_x$ plane. Given a fixed branch choice, there can be a sudden jump of $2\pi\nu_n$ ($\nu_n$ being integer) in the $n$th eigenvalue of $B$ (denoted by $\beta_n$) along certain ``vortex loop" or ``vortex line" in the $k_z$-$k_x$ plane. Then, we need to account for contributions from such $2\pi$ jumps as well, necessitating the vortex-loop contribution, expressed as \cite{Liu-GD-2015}
\begin{equation}
\theta_{\text{VL}}=\sum_n\frac{\phi_n(\mc{C})+\xi_n({\mc{C}})}{2}\nu_n\;,
\label{eq:pre-thetaVL}
\end{equation}
where $n$ labels eigenvalues of $B$. 

Eq.~\eqref{eq:pre-thetaVL} is explained as follows. Suppose $V_{mn}$
diagonalize $B$ such that $B=V\,\beta\, V^\dagger$.
Suppose the $n$th eigenvalue of $B$, denoted as $\beta_n$, has a $2\pi\nu_n$ jump across a vortex loop $\mc{C}$, namely $\beta_{n+}-\beta_{n-}=2\pi \nu_n$, where $+$,$-$ labels two sides of the vortex loop and $\beta_{n+}$ and $\beta_{n-}$ are the limit of $\beta_n$ from the corresponding sides. For the case of single band 3D Chern insulators, we will shown immediately that one can always make a gauge choice such that at one of the eigenvalues $\beta_n$ jumps by $2\pi C$ across an exactly straight line alone $k_z$ direction, as schematically shown in Fig.~\ref{Fig: formalism}(a). In other words, in such a case, $\mc{C}$ is a straight line extending from $k_z=0$ to $k_z=1$ at some given $k_x$. Then, $\phi_n(\mc{C})$ and $\xi_{n}(\mc{C})$ are two kinds of Berry phases calculated along $\mc{C}$: $\phi_n(\mc{C})$ is the Berry phase of $n$th eigenvector (of $B$) $\mathbf{v}_n\equiv (V_{1n}, V_{2n}, \cdots, V_{M n})^{T}$; while $\xi_n(\mc{C})$ is the Berry phase of a unitarily transformed Bloch state $\sum_{m=1}^{M} V_{m n}\ket{u_{m}(k_x,\lambda=0,k_z)}$.

\section{Single-band formalism}
\label{sec:formalism}
\subsection{Gauge choice}
We now start deriving the Chern-Simons OME coupling of a 3D Chern insulator with Chern vector $(0,0,C)$. We note that this is a general situation when discussing bulk properties, since we can always re-define our coordinate system such that the Chern vector $(C_1,C_2,C_3)$ is vanishing in $z$-$x$ and $y$-$z$ plane ~\cite{vanderbilt-berry-book-2018}. We choose the aforementioned gauge with discontinuity at $k_y=0, 1$ plane. We further enforce an additional requirement that in $k_z$, we adopt a periodic parallel-transport gauge (see Appendix $A$), which corresponds to maximally localized hybrid Wannier representation after Fourier transformation with respect to $k_z$ ~\cite{mlwf-prb97}.

It is straightforward to attain our aforementioned 3D gauge by first enforcing a periodic parallel-transport gauge in the $k_x$, then a smooth gauge obtained from parallel transportation in $k_y$ without enforcing periodicity condition, and finally a periodic parallel-transport gauge in $k_z$ \footnote{See Appendix A for detailed explanations of the terminologies such as `periodic parallel-transport gauge' and `parallel transportation' etc.}. The procedures for constructing such a 3D gauge is schematically illustrated in Fig.~\ref{Fig: formalism}(a) and Fig.~\ref{Fig: formalism}(b). Such a gauge construction would push all the potential discontinuity in $k_y$ to the $k_y=1$ plane. Note that enforcing a periodic parallel-transport gauge in $k_z$ will not spoil the smoothness of gauge in $k_x$ and $k_y$ because there is no winding of Berry phase in $k_x$-$k_z$ nor $k_y$-$k_z$ plane due to the corresponding vanishing Chern numbers, so that the gauge transformation itself is smooth in $\mathbf{k}$.

For simplicity, we will first calculate $\theta$ in the case of a single occupied band, where the essential properties of the topological OME response in 3D Chern insulators are already clearly manifested. In the single-band case, $B$ is reduced to a real number $\beta$, and we note that in principle $\beta$ should depend on both $k_z$ and $k_x$. However, we will prove in the following that in the single occupied band case, by insisting a periodic parallel-transport gauge to the Bloch states in the $k_z$ direction, $\beta$ is only dependent on $k_x$. Under such a ``periodic parallel transport gauge" in the $k_z$ direction, the Fourier transformation of the Bloch states with respect to $k_z$ would give rise to the maximally localized Wannier function in $z$ direction.

To see this more explicitly, we first write down the most general gauge-discontinuity condition at the $k_y=0, 1$ plane
\begin{equation}
\ket{u(k_x,k_y=1,k_z)}=e^{i \beta(k_x,k_z)}e^{-i2\pi \hat{y}}\ket{u_n(k_x,k_y=0,k_z)}\;,
\label{eq:cylinder-gauge}
\end{equation}
such that $\beta$ depends on both $k_x$ and $k_z$.

Then, we will show that \(\partial_z \beta=0\) under parallel-transport gauge in $k_z$ direction. Let us define the hybrid Wannier function that is exponentially localized at layer $l$ as 
\begin{equation}
\ket{ W_{ln}(k_x,k_y)}
=\int_0^{1} dk_z \,\vert u_n(k_x,k_y,k_z)\rangle e^{i 2\pi k_z (\hat{z}-l)}
\label{eq:hybrid-wannier-a}
\end{equation}
where $l$ is the layer index (or unit-cell index in $z$ direction). We have chosen the normalization convention that the periodic part of the Bloch function $\ket{u}$ is normalized within the unit cell. For the hybrid Wannier function $\ket{W_{l}(k_x,k_y)}$, it is normalized within the unit cell for the $x$, $y$ coordinates, while normalized over the entire space for the $z$ coordinate.

The inverse transformation of Eq.~\eqref{eq:hybrid-wannier-a} is
\begin{equation}
\ket{u(k_x,k_y,k_z)}=\sum_l e^{i2\pi k_z(l-\hat{z})}\ket{W_{l}(k_x,k_y)}
\label{eq:hybrid-wannier-b}
\end{equation}

Note again, that $ k_z\in[0,1)$ is defined in the reduced coordinate. With periodic parallel-transport gauge in $k_z$, $\ket{ W_l(k_x,k_y)}$ is maximally localized in the $z$ direction, which is the eigenstate of $P\,\hat{z}\,P$ operator ($P$ denoting projection operator to occupied subspace), i.e.,
\begin{equation}
\braket{W_{0}(k_x,k_y)\vert\,\hat{z}\,\vert W_l(k_x,k_y)}
=\bar{z}_0(k_x,k_y)\delta_{0,l}
\label{eq:zbar}
\end{equation}
where \(\bar{z}_0(k_x,k_y)\) is the Wannier center in $z$ direction, which in principle should be dependent on $k_x$ and $k_y$, and may reflect the nontrivial band topology as in the case of three-dimensional topological insulators. However, for weakly coupled Chern-insulator layers, one would expect to get weak $k_x$-$k_y$ dispersions of $\bar{z}_0(k_x,k_y)$.
Multiplying by \(e^{i2\pi k_z'l}\) and summing over \(l\), we have
\begin{equation}
A_z(k_x,k_y,k_z)=2\pi \bar{z}_0(k_x,k_y)
\end{equation}
so that
\begin{subequations}
\begin{align}
&\bar{z}_0(k_x,0)=\frac{i}{2\pi}\braket{u(k_x,0,k_z)|\partial_z u(k_x,0,k_z)}\\ 
&\bar{z}_0(k_x,1)=\frac{i}{2\pi}\braket{u(k_x,1,k_z)|\partial_z u(k_x,1,k_z)}
\label{eq:zbar01}
\end{align}
\end{subequations}
Combining the fact that \(\bar{z}_0(k_x,0)=\bar{z}_0(k_x,1)\) and Eq.~\eqref{eq:hybrid-wannier-b}, Eqs.~\eqref{eq:zbar01}, it follows straightforwardly that
\begin{equation}
\partial_z\beta(k_x,k_z)=0\;.
\end{equation}
This completes our proof that in the single occupied band case under periodic parallel-transport gauge choice in the $k_z$ direction, a smooth but non-periodic gauge in the $k_y$ direction Eq.~\eqref{eq:cylinder-gauge} is always allowed, with the phase factor $\beta$ (characterizing the gauge-discontinuity at the $k_y=0, 1$ plane) being independent of $k_z$. 

Additionally, the nonzero Chern number $C$ within each ($k_x$,$k_y$) plane and periodicity in $k_x$ requires that 
\begin{equation}
\beta(1)-\beta(0)=2\pi C 
\label{single-band-boundary-term-kx}
\end{equation}
if one makes a proper branch choice such that $\beta$ remains continuous from $k_x=0$ to $1$. If a different branch choice is made, there can be a $2\pi C$ jump in $\beta$ at some $k_x^{*}$, which creates a ``vortex line" or ``vortex loop" as explained later.
To see this, one only needs to invoke Stokes' theorem, i.e
\begin{equation}
  \int_0^1dk_x{A_x(k_y=0)}-\int_0^1dk_x{A_x(k_y=1)}=2\pi C
  \label{eq:AxC}
\end{equation}
where we have used the fact that $A_y$ is smooth and periodic along $k_x$ so that the contributions from the other edges cancel. Then Eq.~\eqref{single-band-boundary-term-kx} follows immediately by combining Eq.~\eqref{eq:hybrid-wannier-b} and Eq.~\eqref{eq:AxC}.

\subsection{Derivation of $\theta_{\text{BK}}$ in hybrid Wannier representation}
To put the equation in a more illuminating form, we cast the Berry connection and Berry curvature into the hybrid Wannier representation, the definition of which is given by Eqs.~\eqref{eq:hybrid-wannier-a}-\eqref{eq:hybrid-wannier-b}. 
The Berry connections can be expressed in hybrid-Wannier basis as follows:
\begin{equation}
A_a
=\sum_{l}e^{i2\pi k_zl}A_{a,0l}
\label{eq:AxAy}
\end{equation}
where 
\begin{equation}
A_{a,0l}=\sum_{l}e^{i2\pi k_z l}\,i\braket{W_{0}(k_x,k_y)|\partial_aW_{l}(k_x,k_y)}\nn
\end{equation}
where $a=x,y$. Here we use the convention that \(\ket{u_\k}\) is normalized to 1 over a unit cell and \(\braket{u_{\k}|\partial_x u_{\k}}\) is carried out over a unit cell as well. 
And,
\begin{equation}
A_z=2\pi \bar{z}_0
\label{eq:Az}
\end{equation}
where \(\bar{z_l}(k_x,k_y)=\braket{W_{l}(k_x,k_y)|\hat{z}|W_l(k_x,k_y)}\) is the hybrid Wannier center in the reduced coordinate. It follows from Eq.~\eqref{eq:AxAy} and \eqref{eq:Az} that
\begin{subequations}
\begin{align}
\Omega_{xy} &= \sum_l e^{i2\pi k_z l} \Omega_{xy,0l}
 \label{eq:Oxy} \\
\Omega_{yz}
&= 2\pi \partial_y \bar{z}_0 - i2\pi \sum_l l e^{i2\pi k_z l} A_{y,0l}
\label{eq:Oyz} \\
\Omega_{zx} &= i2\pi \sum_l l e^{i2\pi k_z l} A_{x,0l} - 2\pi \partial_x \bar{z}_0
 \label{eq:Ozx} \\
A_x \Omega_{yz} &= 2\pi \sum_l e^{i2\pi k_z l} A_{x,0l} \partial_y \bar{z}_0 \nn
&\quad - i2\pi \sum_{l', l} e^{i2\pi k_z (l' + l)} l A_{x,0l'} A_{y,0l}
\label{eq:AxOyz} \\
A_y \Omega_{zx} &= i2\pi \sum_{l', l} e^{i2\pi k_z (l' + l)} l A_{y,0l'} A_{x,0l} \nn
&\quad - 2\pi \sum_l e^{i2\pi k_z l} A_{y,0l} \partial_x \bar{z}_0
\label{eq:AyOzx} \\
A_z \Omega_{xy} &= 2\pi \sum_l \bar{z}_0 e^{i2\pi k_z l} \Omega_{xy,0l}
\label{eq:AzOxy}
\end{align}
\end{subequations}
where $\Omega_{xy,0l}=\partial_x A_{y,0l} - \partial_y A_{x,0l}$. Using Eq.~\eqref{eq:pre-thetaBulk}, we put everything together and integrate out the $k_z$ dependence, we arrive at
\begin{align}
\theta_{\rm{BK}}=&-\frac{1}{2}\int dk_x\,dk_y\,\Big(\bar{z}_0\,\Omega_{xy,00}+A_{x,00}\partial_y \bar{z}_0-A_{y,00}\partial_x\bar{z}_0\big)
\label{eq:thetaBK}
\end{align}
where we have used
\begin{equation}
\sum_{l}l(A_{x,l0}A_{y,0l}-A_{y,l0}A_{x,0l})=2\sum_{l}l(A_{x,l0}A_{y,0l})=0
\end{equation}
In the interlayer decoupled limit, \(\bar{z}_0\) is independent of the \((k_x,k_y)\), so the second and the third terms are exactly zeros. For reasons we will see, we define
\begin{align}
  \theta^{\text{anom.}}_{\text{BK}}=&-\frac{1}{2}\int dk_x\,dk_y\,\bar{z}_0\,\Omega_{xy,00}
\end{align}

\subsection{Derivation of $\theta_{\text{GD}}+\theta_{\text{VL}}$ in hybrid Wannier representation}
For \(\theta_{\text{GD}}\), we start with Eq.~\eqref{eq:pre-thetaGD} and the corresponding gauge choice within such a fictitious space are given in Eqs.~\eqref{eq:gauge-disc}. As a reminder, we choose a gauge inside the \(\lambda\)-space that ``linearly interpolates" the gauge discontinuity, i.e,
\begin{equation}
\ket{u(k_x,k_z,\lambda)}=e^{-i\lambda\beta(k_x)}\ket{u(k_x,k_z,\lambda=0)}
\end{equation}
with $\lambda\in{[0,1)}$. We note again that $\ket{u(k_x,k_z,\lambda=0}= \ket{u(k_x,k_y=1,k_z)}$ such that the $\lambda$ linearly interpolates the gauge from $k_y=1$ to that of $k_y=0$. Define as usual
\begin{equation}
\ket{W_{l}(k_x,\lambda)}=\int_{0}^{1}dk_z\,e^{i2\pi k_z(\hat{z}-l)}\ket{u(k_x,\lambda)}
\end{equation}
and
\begin{equation}
\ket{u(k_x,k_z,\lambda)}=\sum_l e^{i2\pi k_z(l-\hat{z})}\ket{W_l(k_x,\lambda)}
\end{equation}
We find the Berry connections can be expressed as
\begin{subequations}
\begin{align}
&A_x(\lambda)=A^0_x+\lambda\, \partial_x\beta\;\\
&A_z=2\pi \bar{z}_0(k_x)\;\\
&A_\lambda=\beta(k_x)
\end{align}
\end{subequations}
where $A_x^0\equiv A_x(\lambda=0)$ and $\bar{z}_0(k_x)\equiv\bar{z}_0(k_x,\lambda)=\langle W_{0}(k_x,\lambda)\vert\hat{z}\vert W_0(k_x,\lambda)\rangle$ is independent of $\lambda$. The Berry curvatures read
\begin{subequations}
\begin{align}
&\Omega_{\lambda z}=0\;\\
&\Omega_{x\lambda}=0\;\\
&\Omega_{zx}=\partial_zA^0_x(k_x,k_z)-2\pi\partial_x\bar{z}_0(k_x)
\end{align}
\end{subequations}
Finally, plugging the above equations into Eq.~\eqref{eq:pre-thetaGD} and taking use of the periodicity in $k_z$, we obtain
\begin{equation}\theta_{\text{GD}}=\frac{1}{2}\int_0^1 dk_x\beta\partial_x\bar{z}_0\end{equation}
For reasons we will soon see, using integration by part, we write it as
\begin{align}
\theta_{\text{GD}}=&\frac{1}{2}\int_0^1dk_x\,\beta\partial_x\bar{z}_0\nn=&\frac{1}{2}(\beta\bar{z}_0)\big|^1_{k_x=0}-\frac{1}{2}\int_0^1\,dk_x\,\bar{z}_0\partial_x\beta
\end{align}
where
\begin{align}
(\beta\bar{z}_0)\big|^1_{k_x=0}
=&\beta(1)\bar{z}_0(k_x=1)-\beta(0)\bar{z}_0(k_x=0)\nn=&(\beta(0)+2\pi C)\bar{z}_0(k_x=0)-\beta(0)\bar{z}_0(k_x=0)\nn=&2\pi C\bar{z}_0(k_x=0)
\end{align}
For the second equality, we have used the fact that the winding of $\beta$ along $k_x$ direction is just $2\pi$ times the Chern number $C$ of the \((k_x,k_y)\) plane and that the $(k_z,k_x)$ plane has Chern number 0 such that $\bar{z}_0(k_x=1)=\bar{z}_0(k_x=0)$. Therefore,
\begin{equation}
\theta_{\text{GD}}=\pi C \bar{z}_0(k_x=0)-\frac{1}{2}\int_0^1\,dk_x\,\bar{z}_0\partial_x\beta(k_x)
\end{equation}
For \(\theta_{\text{VL}}\), since our ``one dimensional matrix" $\beta$ is always diagonalized, we are free to choose the ``eigenvector" of $\beta$ (which only has one element), $V_{11}=1$, which certainly has zero Berry phase. This eliminates the first term in Eq.~\eqref{eq:pre-thetaVL}, leaving us with
\begin{equation}
\theta_{\text{VL}}=-\frac{1}{2}C\int_{0}^1 A^0_z(k_x=0)dk_z=-\pi C\bar{z}_0(k_x=0)\;,
\end{equation}
where we have used $\bar{z}_0=\frac{1}{2\pi}\xi$, with $\xi$ denoting the Berry phase of the physical Bloch state winding around the vortex loop, as explained near Eq.~\eqref{eq:pre-thetaVL}. The minus sign comes from the fact that the discontinuity of $\beta$ is manifested as a local $-2\pi C$ shift across the vortex loop (or vortex line) at $k_x^{*}$ (see Fig.~\ref{Fig: formalism}(c)).
Combining the two contributions, we arrive at
\begin{equation}
\theta_{\text{GD}}+\theta_{\text{VL}}=-\frac{1}{2}\int_0^1\,dk_x\,\bar{z}_0(k_x=0,\lambda=0) \partial_x\beta \equiv\theta^{\text{anom.}}_{\text{GDVL}}
\label{eq:theta-GDVL}
\end{equation}
We see now the boundary term for $\theta_{\text{GD}}$ is canceled by a term in $\theta_{\text{VL}}$. Here we call this term the ``anomalous" Chern-Simons term, denoted by $\theta^{\text{anom.}}_{\text{GDVL}}$. The name ``anomalous" will be justified later.

\section{Extension to multiple occupied bands}
\label{sec:multi-band}
We have only considered a single occupied band in the above derivations. The entire formalism also applies to the situation of $M$ occupied bands. As a reminder, now $B_{mn}(k_x,k_z)$ is $M\times M$ Hermitian matrix element at each $(k_x,k_z)$ defined in occupied-band basis, the eigenvalues of which $\beta_n(k_x,k_z)$ ($n=1,...,M$) satisfy
\begin{equation}
\sum_{n=1}^{M}\,(\,\beta_n(k_x=1)-\beta_n(k_x=0)\,)=2\pi C\;.
\end{equation}
Again, a periodic parallel transport gauge in the $k_z$ direction is assumed. Note, we have restored the dependence of $B$ on $k_z$, since a similar analysis will show. 
\begin{equation}
\partial_z B_{mn}=-\sum_{r}\bar{z}_re^{iB_{mr}}e^{-iB_{rn}}+\bar{z}_n\delta_{mn}
\end{equation}
In other words, only the trace of $B$ is constant along $k_z$. In the rest of this section, we will show the main results and leave the derivations to Appendix B.

\subsection{Derivation of $\theta_{\text{BK}}$ in hybrid Wannier representation}
With these extensions, Eq.~\eqref{eq:pre-thetaGD} becomes
\begin{align}
\theta_{\text{BK}}=&-\frac{1}{2}\,\int dk_xdk_y\Big(\,\sum_n z_{0n}{\Omega_{xy,0n,0n}}\;\nn&+\sum_m(A_{x,0n,0n}\partial_{y}\bar{z}_{0n}-A_{y,0n,0n}\partial_{x}\bar{z}_{0n})\;\nn
&+2\sum_{lmn}ilA_{x,lm,0n}A_{y,0n,lm} \;\nn&-2i\sum_{lmn}(\bar{z}_{0m}-\bar{z}_{0n})A_{x,lm,0n}A_{y,0n,lm}\,\Big)
\label{eq:thetaBK-multi}
\end{align}
where 
\begin{subequations}
\begin{equation}
A_{a,0n,lm}=i\langle W_{0,n}(k_x,k_y)\vert\partial_a W_{l,m}(k_x,k_y)\rangle
\end{equation}
\begin{equation}
\Omega_{xy,0n,lm}=\partial_xA_{y,0n,lm}-\partial_y A_{x,0n,lm}
\end{equation}
\end{subequations}
Here, $\ket{W_{l,m}(k_x,k_y)}$ denotes the $m$th hybrid Wannier function localized in layer $l$ and $\bar{z}_{0n}=\langle W_{0,n}(k_x,k_y)\vert \hat{z}\vert W_{0,n}(k_x,k_y)\rangle$. 
We note the result resembles that presented in Ref.~\onlinecite{Maryam-2015}, the main difference being the extra $\frac{1}{2}$ on the term related to $\Omega_{xy,00,00}$ and $z_{00}$. 
We will see that we will be saved by the discontinuity contributing a similar term. 
Even in the multi-band cases, with non-vanishing interlayer coupling, the bulk term may vanish due to constraint from crystalline symmetries such as inversion symmetry. In the presence of inversion symmetry, it is straightforward to show that
\begin{subequations}
\begin{align}
&A_a(\k)=-A_a(-\k)\;\\
&\Omega_{ab}(\k)=\Omega_{ab}(-\k)
\end{align}
\end{subequations}
Then, clearly Eq.~\eqref{eq:thetaBK-multi} vanishes up to $\pi C$ (see Sec.~\ref{sec:quantization}).

\subsection{The Expression for $\theta_{\text{GD}}+\theta_{\text{VL}}$}
Following Eq.~\eqref{eq:pre-thetaGD} and Eq.~\eqref{eq:pre-thetaVL}, we are able to find the gauge-discontinuity and vortex-loop contributions can be expressed as
\begin{align}
\theta_{\textrm{GD}}+\theta_{\textrm{VL}}=&-\frac{1}{4\pi}\int dk_xd\lambda dk_z\, \mathrm{Tr}\,(-iB[\partial_zM,\partial_xM^\dagger]\nn&+iM[\partial_zB,\partial_x M^\dagger]+iM[\partial_zM^\dagger,\partial_xB]\nn&+2B[\partial_z M^\dagger,MA_x^0]+\partial_zB[M^\dagger,MA_x^0]\nn&+B\partial_zA_x^0-2B[\partial_x M^\dagger,MA_z^0]\nn&-\partial_xB[M^\dagger,MA_z^0])\nn&+\int_0^1dk_z\,\sum_{mn}V^\dagger_{mn}\partial_zV_{nm}C_n+\theta^{\text{anom.}}_{\text{GDVL}}
\end{align}
where 
\begin{equation}
\theta^{\text{anom.}}_{\text{GDVL}}=-\frac{1}{4\pi}\int dk_zdk_z\,\mathrm{Tr}(\partial_xB\,A^0_z)
\end{equation}

\begin{figure*}[hbt]
  \centering
  \includegraphics[width=\linewidth]{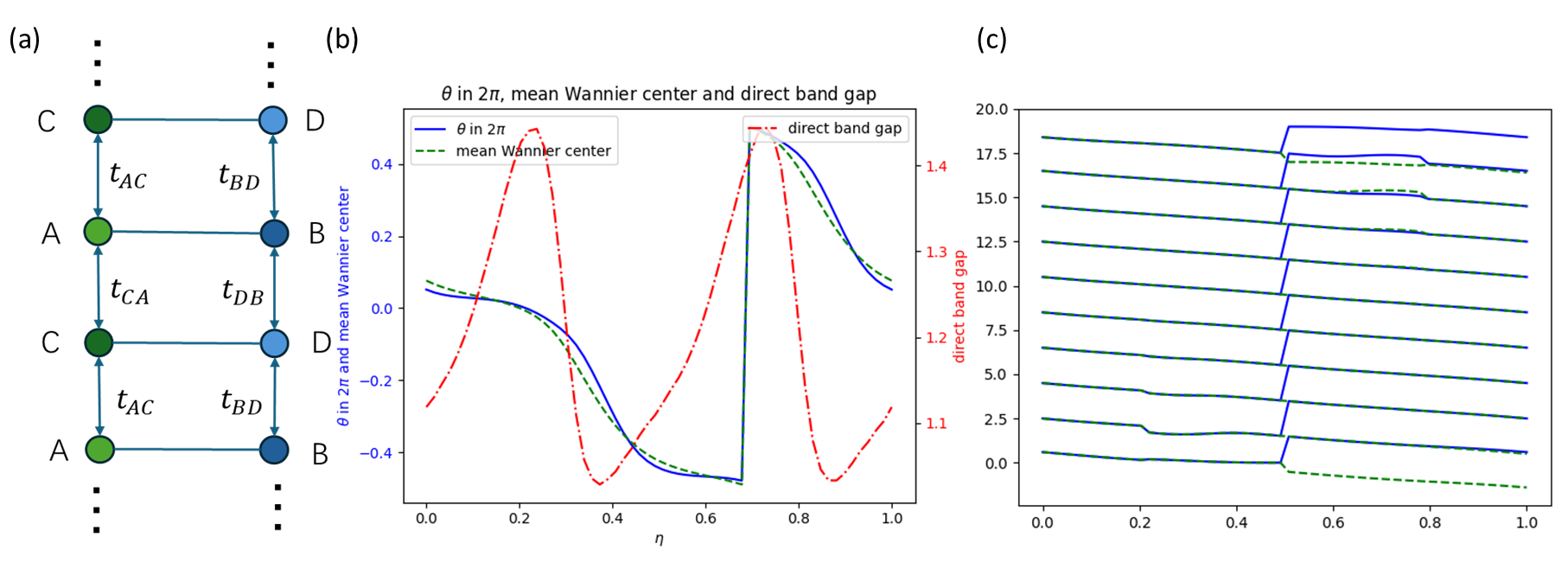}
  \caption{(a) A schematic illustration of the side view $AA$-stacked Haldane model with two layers (four sublattices) in each unit cell. (b) $\theta$-pumping of the stacked Haldane model. (c) Wannier centers of 10-unit-cell slab at high symmetry point $M$ plotted as a function of pumping parameter $\eta$. The solid blue line denotes hybrid Wannier centers $\bar{z}$ satisfying periodicity $\bar{z}(\eta=0)=\bar{z}(\eta=1)$, and the green dashed line denotes Wannier centers shifted down by 2 atomic layers after the jump such that it evolves mostly continuously with $\eta$.}
  \label{Fig: pumping}
\end{figure*}

\section{Quantized difference of layer Chern-Simons coupling}
\label{sec:quantization}
Despite that we have enforced a periodic parallel transport gauge, the formalism still admits a ``purely radical'' gauge transformation, 
\begin{equation}
\ket{u_{n\mathbf{k}}}\mapsto e^{-i2\pi p k_z}\ket{u_{n\mathbf{k}}}\;.
\label{eq:gauge-multi}
\end{equation}
Because a linear-in-$k_z$ phase factor is added, the hybrid Wannier function is still maximally localized in $z$ after such gauge transformation; the only effect is to re-label the Wannier center, $\bar{z}_{0n}\to \bar{z}_{0n}+p$, leaving the shapes of the hybrid Wannier functions unchanged.
In the single-band case, it is straightforward to show that
\begin{align}
\int dk_x\,dk_y\,\Omega_{xy,00}=2\pi C\;, 
\end{align}
which is clearly gauge invariant. Therefore, under the gauge transformation of Eq.~\eqref{eq:gauge-multi},
\begin{equation}
\theta^{\text{anom.}}_{\text{BK}}\equiv-\frac{1}{2}\int dk_x\,dk_y\,\bar{z}_0\Omega_{xy,00}\mapsto\theta^{\text{anom.}}_{\text{BK}}-\pi pC
\label{eq:thetaBK-pump}
\end{equation}
with other terms being gauge invariant. Likewise,
\begin{equation}
\theta^{\text{anom}.}_{\text{GDVL}}\equiv-\frac{1}{2}\int_0^1\bar{z}_0(0)\partial_x\beta dk_x\mapsto\theta^{\text{anom.}}_{\text{GDVL}}-\pi pC
\label{eq:thetaGDVL-pump}
\end{equation}
The other terms are all gauge invariant in the sense of Eq.~\eqref{eq:gauge-multi}. This means, by going from $l$th to $(l+p)$th layer, $\theta^{\text{anom.}}=\theta^{\text{anom.}}_{\text{BK}}+\theta^{\text{anom.}}_{\text{GDVL}}$ changes by $-2\pi p C $, as schematically illustrated in Fig.\ref{Fig: formalism}(c). This ambiguity, unique to 3D Chern insulators, originates from the ambiguity in defining hybrid Wannier center in $z$ direction, multiplied by the Chern number defined in $(k_x, k_y)$ plane. Physically, it is interpreted as a kind of quantized layer-resolved OME coupling that only appears along the vertical stacking direction (or, along the direction of the Chern vector). The total response from all layers may still vanish due to the constraint from crystalline symmetries, but each layer would have nontrivial OME response in such a way that the difference between the responses of adjacent layers is exactly $-C e^2/h$ \cite{lu_2024}, 
\begin{equation}
\alpha_{zz}^{(l+1)}-\alpha_{zz}^{(l)}=-Ce^2/h\;. 
\label{eq:quantize-alpha}
\end{equation}
Or, equivalently, the spatial gradient of $\alpha$ in $z$ is quantized as 
\begin{equation}
\nabla_z \alpha =-C \frac{e^2}{h\,d_0} \;,
\end{equation}
where $d_0$ is the lattice constant in $z$ direction.
The above proof shows that this quantization rule is exact in the presence of arbitrary interlayer coupling as long as the bulk gap remains opened up. It is expected to be robust against disorder, as can be inferred from numerical studies and argument based on Středa's formula \cite{lu_2024,streda-1982}. 

In the multi-band case, the same can be said with
\begin{align}
\theta^{\text{anom.}}=&-\frac{1}{2}\,\int dk_xdk_y\sum_{n=1}^{M} \bar{z}_{0n}(k_x,k_y)\Omega_{xy,0n,0n}(k_x,k_y)
\nn&-\frac{1}{4\pi}\int dk_xdk_z\,\mathrm{Tr}(\partial_xBA_z^0)
\end{align}
where we use the following equations 
\begin{subequations}
\begin{align}
&\int dk_x dk_y \sum_{n=1}^{M}\Omega_{xy,0n,0n}=2\pi C\;\\
&\mr{Tr}(B)\bigg|^{1}_{k_x=0}=2\pi C
\end{align}
\end{subequations}
It can also be verified that the rest of the terms are invariant, by noting under the gauge transformation, $A_x$, $B$ are unchanged. Therefore, all the discussions in the single-band case still applies to the situation of multiple occupied bands.

\section{Adiabatic pumping}
\label{sec:pumping}
\subsection{$\theta$-pumping in a stacked Haldane model}
The ambiguity of $2\pi C$ in defining layer-resolved $\theta$ suggests a possible cyclic adiabatic pumping process through which the OME coupling projected to each layer may be changed by exactly $Ce^2/h$, corresponding to a quantized pumping of Chern-Simons coupling from bottom to top layers. Interestingly, according to Eqs.~\eqref{eq:gauge-multi}-\eqref{eq:thetaGDVL-pump} and the discussions therein, the pumping of $\theta$ (if there is any) would be directly associated with the pumping of Wannier center in the $z$ direction. Here we apply our formalism to $AA$ (head-to-head) stacked Haldane model, and directly illustrate such a pumping process. 
 In the $x$-$y$ plane the hoppings of the model are identical to monolayer Haldane model \cite{Haldane-model}, including nearest-neighbor (NN) and next-nearest-neighbor (NNN) hoppings, allowing us to manipulate the Chern number of the $k_x$-$k_y$ plane. As illustrated in Fig.~\ref{Fig: pumping}(a), we have 4 atoms in a unit cell of two layers, denoted by $A$, $B$, $C$, $D$ sublattices, with NN interlayer hoppings $t_{AC}$ and $t_{BD}$. We consider the minimal case of only one occupied band. The Wannier centers are pinned at atomic layers or the midplanes between two layers for model parameters with mirror symmetry. The pumping process involves evolving the interlayer coupling $t_{AC/CA}$ (and/or $t_{BD/DB}$) and onsite energy $\delta_{\alpha}$ ($\alpha=A, B, C, D$). In order to allow for non-vanishing dispersion of Wannier centers, we let $-A-C-A-$ pumping and $-B-D-B-$ pumping to differ. The hopping terms are chosen to depend on adiabatic parameter $\eta\in [0,1)$ in the following way:
\begin{align}
&\delta_A=1.5\cos(2\pi\eta+\pi/2)\;\nn
&t_{AC}=1+\cos(2\pi\eta)\;\nn
&\delta_C=1.5\cos(2\pi\eta+3\pi/2)\;\nn  
&t_{CA}=1+\cos(2\pi\eta+\pi)\;\nn
&\delta_B=3\cos(2\pi\eta+\pi/2+2/5\pi)\;\nn
&t_{BD}=0.5+0.5\cos(2\pi\eta+2/5\pi)\;\nn
&\delta_D=3\cos(2\pi\eta+3\pi/2+2/5\pi)\;\nn  
&t_{DB}=0.5+0.5\cos(2\pi\eta+\pi+2/5\pi)\;,
\label{eq:pumping}
\end{align}
where the $t$s are interlayer hoppings and the $\delta$s are onsite energies, as explained above. The in-plane hoppings remain constant with NN hopping and NNN hopping being $0.8$ and $0.3i$ respectively. With the above choice of parameters, we confirm that the lowest-energy band carriers a Chern number $C=-1$ within $(k_x,k_y)$ plane. Moreover, the direct bulk gap is not closed throughout the adiabatic pumping process as shown by the red dashed line in Fig.~\ref{Fig: pumping}(b), so that the Chern number is unchanged.
As shown by the solid blue line in Fig.~\ref{Fig: pumping}(b), given a branch choice $\theta\in[-\pi,\pi)$, there is a $2\pi$ jump in bulk $\theta$ (calculated using the gauge-discontinuity formalism introduced above) through the adiabatic pumping process parameterized by Eqs.~\eqref{eq:pumping}. One can also make a proper branch choice such that $\theta$ evolves continuously through the cyclic pumping process, then $\theta$ would be changed by exactly $2\pi C$ through the cyclic adiabatic evolution. This indicates that such pumping process is associated with a non-trivial second Chern number. We also inspect the behavior of hybrid Wannier center $\bar{z}$ (averaged over $k_x$ and $k_y$), and we see exactly the same $2\pi$ jump during the above cyclic pumping process as shown by green dashed line in Fig.~\ref{Fig: pumping}(b). This clearly demonstrates that the quantized shift of $\theta$ is directly associated with that of hybrid Wannier center.
In the case of the $\theta$-pumping of an insulator with vanishing Chern vector as proposed in Ref.~\onlinecite{Maryam-2015}, multiple occupied bands must be involved, and a nontrivial pumping process is necessarily accompanied with touching events between different hybrid Wannier center sheets during the cyclic evolution in order to exchange $2\pi$ quanta of Berry flux. This can be traced back to the fact that single-band $\theta$ is well defined without any $2\pi$ ambiguity if the Chern number vanishes \cite{vanderbilt-berry-book-2018}. Moreover, the ``$\theta$-pumping" process reported in Ref.~\onlinecite{Maryam-2015} is not associated with any charge pumping. In our case of 3D Chern insulators, the story is different: the pumping of $\theta$ is tied to the pumping of exactly one charge (per unit cell) from one unit cell to its neighboring cell (or, from top to bottom surfaces in slab geometry), and such pumping can be realized with only one occupied band without the necessity of any touching event between Wannier center sheets thanks to the nonzero Chern number. Such $\theta$ pumping in the case of single occupied band is all attributed to the anomalous Chern-Simons term discussed above, which is unique to 3D Chern insulators. After the cycle, an integer quantum of $-C e^2/h$ of layer-resolved Chern-Simons coupling would be pumped from one unit cell to its neighboring cell, concomitant with the pumping of one charge quantum along the direction of the Chern vector. Heuristically, the anomalous contribution to the $\theta$ term may be interpreted as the Berry curvature within the plane of non-zero Chern number (i.e., $(k_x, k_y)$ plane) coupled with the Wannier center in the vertical ($z$) direction. Thus, any quantized pumping of Wannier center would ``carry" the integrated Berry curvature $2\pi C$ from one unit cell to the other.
 
We also consider the pumping of Wannier charges in a slab geometry (with open boundary condition in $z$) for better understanding of the adiabatic cyclic process. For simplicity, this time we let $-A-C-A-$ pumping and $-B-D-B-$ pumping to coincide, i.e., with $t_{AC}=t_{BD}=1+\cos(2\pi\eta)$, $\delta_A=\delta_B=1.5\cos(2\pi\eta+\frac{\pi}{2})$ , $t_{CA}=t_{DB}=1+\cos(2\pi\eta+\pi)$ and $\delta_{C}=\delta_{D}=1.5\cos(2\pi\eta+\frac{3\pi}{2})$. For clarity's sake, we turn off the in-plane hoppings of the very top and the very bottom atomic layers so that we have flat surface bands. We have also fine tuned the onsite energy of the two surface layers (without changing any bulk property) so that the two surface bands move through each other in the bulk energy gap. 
With our parameters, the occupied band of the $AA$-stacked Haldane model has the Chern vector $(0,0,-1)$ throughout the pumping process. To be more specific, we consider a 10-unit-cell slab and calculate the evolution of Wannier centers $\bar{z}_l$ ($1\leq l\leq 10$) as a function the pumping parameter $\eta$. As shown by the solid blue lines in Fig.~\ref{Fig: pumping}(c), each Wannier center has a jump of 2 (which is the lattice vector in $z$ direction) when $\eta$ is around 0.5, in order to restore periodicity after the cyclic adiabatic evolution, i.e., $\bar{z}(\eta=0)=\bar{z}(\eta=1)$. However, physically the Wannier center has to evolve continuously with the pumping parameter $\eta$, and it is well defined modulo 2 (the vertical lattice constant). Therefore, in order to demonstrate the physical pumping process, the Wannier center of each unit cell is shifted down by 2 atomic layers (one unit cell) after the jump such that $\bar{z}$ evolves (mostly) continuously with $\eta$, as shown by the green dashed line in Fig.~\ref{Fig: pumping}(c). Clearly, the Wannier centers are transported toward negative $z$ direction by an exact integer quanta (one unit cell) after a cyclic adiabatic evolution,
which precisely corresponds to the pumping of one charge (per unit cell) from top to bottom surface layers. As discussed above, such a quantized charge pumping is always accompanied by the pumping of quantized Chern-Simons coupling. 


\section{Summary}
\label{sec:summary}

To summarize, using the gauge-discontinuity formalism, we have derived a general formula of Chern-Simons OME coupling for 3D Chern insulators with nonzero Chern number $C$ within the $(k_x, k_y)$ plane, characterized by a Chern vector $(0,0,C)$. When expressed in hybrid Wannier basis, we demonstrate that there is an anomalous contribution to the Chern-Simons coupling, which is well defined up to $-C e^2/h$. Physically, such anomalous term results in a quantized difference $-C e^2/h$ between Chern-Simons couplings projected to two adjacent layers, as given by Eq.~\eqref{eq:quantize-alpha}. Such a result is mathematically exact for 3D Chern insulators with arbitrary interlayer couplings, as long as the bulk gap is is not closed. This result also applies to the case of 3D Chern insulators with an arbitrary Chern vector $(C_1, C_2, C_3)$, as if one is only interested in the bulk properties, one can always rotate the coordinate system such that the Chern vector in the rotated frame only has one non-vanishing component (defined as the $z$ component). Interestingly, we find that the anomalous contribution of Chern-Simons coupling may be heuristically interpreted as Berry curvature of occupied bands ``coupled" with its hybrid Wannier center in the direction of Chern vector. As a result, one can carefully design an cyclic adiabatic process through which the Wannier center is pumped from one unit cell to the neighboring cell by an exact integer quantum, this would lead to the integral pumping of layer-projected Chern-Simons coupling by a quantized value $-C e^2/h$. We explicitly demonstrate such quantized pumping process of Chern-Simons coupling in a stacked multilayer Haldane model. Lastly, we note that the quantized spatial gradient of Chern-Simons coupling derived in this work may fundamentally change the electrodynamics \cite{axion-ed} in a (so far hypothetical) medium consisted of 3D Chern insulators. We hope that our work would motivate further theoretical studies and materials search of 3D Chern insulators. 
Our work also opens avenues for designing topological quantum phenomena in layered systems.

\acknowledgements 
This work is supported by National Key Research and Development Program of China (grant No. 2024YFA1410400 and No. 2020YFA0309601) and the National Natural Science Foundation of China (grant No. 12174257).

\appendix

\section{Parallel transport gauges}
In this section, we will recall the definitions of periodic parallel transport gauge and how to enforce it numerically.

In the continuous case, defined on a path of the Brillouin zone, a transport gauge means the matrix $A_{\alpha}$ is vanishing on the path. In the discrete case, a parallel transport gauge on a path is defined by making sure the overlap matrices between two neighboring $\k$ points 
\begin{equation}
  \mathcal{M}^{{\mathbf{k}_i},\mathbf{k}_{i+1}}_{mn}=\braket{u_m(\mathbf{k}_i)|u_n(\mathbf{k}_{i+1})}
\end{equation}
are Hermitian~\cite{vanderbilt-berry-book-2018}, which can be implemented using singular value decomposition (SVD). 

For concreteness, suppose we have numerically obtained $\ket{u_n(\mathbf{k}_{i})}$ whose overall phase are not in our control. We start at states at $\mathbf{k}_0$ and then calculate its overlap with $\mathbf{k}_1$, which is not not necessarily Hermitian. Under SVD, 
\begin{equation}
  \mathcal{M}^{0,1}=V\Sigma W^\dagger
\end{equation}
where $\Sigma$ is diagonal and V and W are unitary. Next we apply $VW^\dagger$ to $\ket{u_n(\mathbf{k}_1)}$. It can be checked that the new overlap matrix is Hermitian. In this way, we can align the overlap matrix between all $\mathbf{k}'s$ sequentially.
The gauge can always be made with the caveat of a mismatch on the start and the end if it's enforced on the loop. This mismatch is characterized by the overlap matrix between the end states and the start states after we enforce the above procedure. It's straightforward to see its eigenvalues $\phi_n$ is nothing but the multi-band Berry phases on the loop. We can get rid of the mismatch by rotating the states to a ``periodic parallel transport gauge". That is,
\begin{equation}
  \ket{u'_{n}(\mathbf{k}_j)}=e^{i\phi_n(j/N)}\ket{u_{n}(\mathbf{k}_j)}
\end{equation}
where a regular parallel transport gauge has already been enforced on $\ket{u_{n}(\mathbf{k}_j)}$'s and $\phi_n$ is the Berry phase along the loop. In other words, the winding is uniformly distributed to each $\mathbf{k}$. 

\section{Derivation of multi-band $\theta_{\text{GD}}+\theta_{\text{BK}}$}
Due to the non-Abelian nature in the multi-band case, there's considerable more algebra than the single band case. Nevertheless, we find again using Eq.~\eqref{eq:pre-thetaBulk}
\begin{subequations}
\begin{flalign}
&\begin{aligned}
A_x &= iW\partial_xW^\dagger + WA_x^0W^\dagger
\end{aligned}&&\\
&\begin{aligned}
A_\lambda &= B
\end{aligned}&&\\
&\begin{aligned}
A_z &= iW\partial_zW^\dagger + WA_z^0W^\dagger
\end{aligned}&&\\
&\begin{aligned}
\Omega_{\lambda z} &= -\partial_zB + W\partial_zBW^\dagger + iBWA_z^0W^\dagger \\
&\quad - iWA^0_zBW^\dagger
\end{aligned}&&\\
&\begin{aligned}
\Omega_{zx} &= i\partial_zW\partial_xW^\dagger + \partial_zWA_x^0M^\dagger + W\partial_zA_x^0W^\dagger \\
&\quad + WA_x^0\partial_zW^\dagger - i\partial_xW\partial_zW^\dagger - \partial_xWA_z^0W^\dagger \\
&\quad - W\partial_xA_z^0W^\dagger - WA_z^0\partial_xW^\dagger
\end{aligned}&&\\
&\begin{aligned}
\Omega_{x\lambda} &= \partial_xB - W\partial_xBw^\dagger - iBWA_x^0W^\dagger \\
&\quad + iWA^0_xBW^\dagger
\end{aligned}&&\\
&\begin{aligned}
[A_z, A_x] &= [WA^0_zW^\dagger, WA^0_xW^\dagger] + i[W\partial_zW^\dagger, WA^0_xW^\dagger] \\
&\quad + i[WA_z^0W^\dagger, W\partial_xW^\dagger] - [W\partial_zW^\dagger, W\partial_xW^\dagger]
\end{aligned}&&
\end{flalign}
\end{subequations}
where $A_{z,mn}^0=\braket{u_m(\lambda=0)|\partial_zu_n(\lambda=0)}$. The trace evaluates to 
\begin{align}
&\mathrm{Tr}\,( A_z\Omega_{x\lambda}+A_x\Omega_{\lambda z}+A_\lambda\Omega_{zx}-2i [ A_z, A_x ] A_\lambda)\nonumber\nn
=&\mathrm{Tr}\,(-iB[\partial_zW,\partial_xW^\dagger]+iW[\partial_zB,\partial_x W^\dagger]\nn&+iW[\partial_zW^\dagger,\partial_xB]+2B[\partial_z W^\dagger,WA_x^0]+\partial_zB[W^\dagger,WA_x^0]\nn&+B\partial_zA_x^0-2B[\partial_x W^\dagger,WA_z^0]-\partial_xB[W^\dagger,WA_z^0]\nn&-B\partial_xA_z^0)
\end{align}
so that,
\begin{align}
\theta_{\textrm{GD}}=&-\frac{1}{4\pi}\int dk_xd\lambda dk_z\, \mathrm{Tr}\,(-iB[\partial_zM,\partial_xM^\dagger]\nn&+iM[\partial_zB,\partial_x M^\dagger]+iM[\partial_zM^\dagger,\partial_xB]\nn&+2B[\partial_z M^\dagger,MA_x^0]+\partial_zB[M^\dagger,MA_x^0]\nn&+B\partial_zA_x^0-2B[\partial_x M^\dagger,MA_z^0]\nn&-\partial_xB[M^\dagger,MA_z^0])+\frac{1}{4\pi}\int dk_xdk_z\mathrm{Tr}\,(B\partial_xA_z^0)
\end{align}
The last term again can be written as 
\begin{align}
&\frac{1}{4\pi}\int dk_z\mathrm{Tr}\bigg(\int dk_x\,B\partial_xA_z^0\bigg)\nonumber\\
=&\frac{1}{4\pi}\int dk_z\,\mathrm{Tr}(BA_z^0)\bigg|_{k_x=0}^1-\frac{1}{4\pi}\int dk_zdk_z\,\mathrm{Tr}(\partial_xBA_z^0)\nonumber\\
=&\frac{1}{2}\int dk_z\,\sum_{n}B_{nn}\bigg|_{k_x=0}^1\overline{z}_n(k_x=0,\lambda=0)\nn&-\frac{1}{4\pi}\int dk_zdk_x\,\mathrm{Tr}(\partial_xBA_z^0)
\label{eq:thetaGD-anom}
\end{align}
In the last equality we've used the fact that $A_z^0$ is diagonal. For the $\xi(\mc{C})$ term of $\theta_{\text{VL}}$ , we have
\begin{align}
&-\frac{i}{2}\sum_{mnr}\int_0^1 dk_z \,V^\dagger_{mr}V_{nm} C_m\braket{u_{r}(0,k_z,0)|\partial_z|u_{n}(0,k_z,0)}\nn
=&-\frac{1}{4\pi}\sum_{nr}\int_0^1 dk_zA^0_{z,nn}\delta_{rn} (B_{nr}(k_x=1)-B_{nr}(k_x=0))\nn
=&-\frac{1}{2}\sum_{n}\int_0^1 dk_zB_{nn}\bigg|_{k_x=0}^1\bar{z}_n(k_x=0,\lambda=0)
\end{align}
When deriving the first equality, we have used the fact that $\beta_m(k_x=1)-\beta_m(k_x=0)=2\pi C_m$, and $\beta_m$ is the $m$th eigenvalue of $B$ matrix, thus satisfying $\sum_{m}V_{nm}\beta_m V_{mn}^{\dagger}=B_{nn}$. This term exactly cancels the first term of Eq.~\eqref{eq:thetaGD-anom}. The $\phi(\mc{C})$ term is
\begin{equation}
\int_0^1dk_z\,\sum_{mn}V^\dagger_{mn}\partial_zV_{nm}C_n
\end{equation}
We define the $\theta^{\text{anom.}}_{\text{GDVL}}$ in the multi-band case as
\begin{equation}
\theta^{\text{anom.}}_{\text{GDVL}}=-\frac{1}{4\pi}\int dk_zdk_z\,\mathrm{Tr}(\partial_xBA^0_z)
\end{equation}
\bibliography{reference}

\end{document}